\documentclass[a4paper,11pt]{article}
\pdfoutput=1
\usepackage{jcappub}
\usepackage{hyperref}
\hypersetup{colorlinks,urlcolor=blue,citecolor=black,linkcolor=black,filecolor=black}
\usepackage{orcidlink}
\usepackage{color}
\usepackage{setspace}
\usepackage[T1]{fontenc}
\title{\boldmath Time-Machines Construct in $f(\mathcal{R},\mathcal{A},A^{\mu\nu}\,A_{\mu\nu})$ and $f(\mathcal{R})$ Modified Gravity Theories}

\author[a]{F. Ahmed\orcidlink{0000-0003-2196-9622},}
\author[b]{J. C. R. de Souza\orcidlink{0000-0002-7684-9540},}
\author[b,1]{and A. F. Santos\orcidlink{0000-0002-2505-5273}\note{Corresponding author}}

\affiliation[a]{Department of Physics, University of Science \& Technology Meghalaya, Ri-Bhoi, Meghalaya, 793101, India}
\affiliation[b]{Programa de P\'{o}s-Gradua\c{c}\~{a}o em F\'{\i}sica, Instituto de F\'{\i}sica, Universidade Federal de Mato Grosso, Cuiab\'{a}, Brasil}

\emailAdd{faizuddinahmed15@gmail.com}
\emailAdd{jean.carlos@fisica.ufmt.br}
\emailAdd{alesandroferreira@fisica.ufmt.br}

\vspace{0.3cm}

\abstract{In this paper, our objective is to explore a time-machine space-time formulated in general relativity, as introduced by Li (Phys. Rev. D {\bf 59}, 084016 (1999)), within the context of modified gravity theories. We consider Ricci-inverse gravity of all Classes of models, {\it i.e.}, (i) Class-{\bf I}: $f(\mathcal{R}, \mathcal{A})=(\mathcal{R}+{\kappa\,\mathcal{R}^2}+\beta\,\mathcal{A})$, (ii) Class-{\bf II}: $f(\mathcal{R}, A^{\mu\nu}\,A_{\mu\nu})=(\mathcal{R}+{\kappa\,\mathcal{R}^2}+\gamma\,A^{\mu\nu}\,A_{\mu\nu})$ model, and (iii) Class-{\bf III}: $f(\mathcal{R}, \mathcal{A}, A^{\mu\nu}\,A_{\mu\nu})=(\mathcal{R}+{\kappa\,\mathcal{R}^2}+\beta\,\mathcal{A}+\delta\,\mathcal{A}^2+\gamma\,A^{\mu\nu}\,A_{\mu\nu})$ model, where $A^{\mu\nu}$ is the anti-curvature tensor, the reciprocal of the Ricci tensor, $R_{\mu\nu}$, $\mathcal{A}=g_{\mu\nu}\,A^{\mu\nu}$ is its scalar, and $\beta, {\kappa}, \gamma, \delta$ are the coupling constants. Moreover, we consider $f(\mathcal{R})$ modified gravity theory and investigate the same time-machine space-time. In fact, we show that Li time-machine space-time serve as valid solutions  both in Ricci-inverse and $f(\mathcal{R})$ modified gravity theories. Thus, both theory allows the formation of closed time-like curves analogue to general relativity,  thereby representing a possible time-machine model  in these gravity theories theoretically.
}


\arxivnumber{2407.11513}

\begin{document}
\maketitle
\flushbottom

\section{Introduction}

The physical theories, such as the Standard Model and General Relativity, are constructed with the assumption that causality is a characteristic obeyed by the system. In physics, causality ensures that an effect cannot occur before its cause, preserving the logical sequence of events in time. Hawking proposed the chronology protection conjecture, asserting that physical laws prevent the violation of causality \cite{Hawking}. However, numerous counterexamples to this conjecture have been identified. General Relativity, the most successful theory of gravity, permits certain exact solutions that result in causality violations. One of the best-known solutions with this characteristic was proposed by Kurt G\"{o}del in 1949 \cite{Godel}. The G\"{o}del solution is the first exact solution of General Relativity with rotating matter. The primary characteristic of this solution is the presence of Closed Time-like Curves (CTCs), which enable a traveler following this trajectory to travel back in time. Consequently, CTCs function as theoretical time machines, leading to a violation of causality. G\"{o}del-type solutions are introduced to explore universes where causality can be violated \cite{tipo_godel}. In this context, these solutions allow for the construction of both causal and non-causal scenarios depending on the universe's matter content. The violation of causality from exact solutions of general relativity is not limited to Gödel or Gödel-type solutions. Other models also demonstrate this phenomenon, such as Kerr and Kerr-Newman black holes \cite{ctc2, ctc3}, Lanczos-van Stockum model \cite{Van,CL}, cosmic strings \cite{ctc1}, Tipler rotating cylinder \cite{Tip},  Bonnor and Steadman space-time \cite{Bon}, the Ori time-machine metric \cite{Ori}, and several others detailed in \cite{FA1, FA22, FA3, FA4}. Another important solution to highlight here is the one proposed by Gott and Li \cite{GoLi}. In this model, CTCs could play a crucial role in the early universe. If we trace back the history of time, we might encounter an early epoch where CTCs exist, implying that there is no earliest event in time. The primary objective of this research is to explore a time-machine space-time formulated in the framework of general relativity theory, as originally proposed by Li \cite{LX} (see, also \cite{LX2}). This space-time serves as a time machine model, manifesting as closed time-like curves within the temporal region of the geometry. Additionally, Li's work demonstrates that a Misner-like anti-de Sitter space can be derived from the covering space of anti-de Sitter space. This Misner-like space contains closed time-like curves (CTCs), but the regions with CTCs are separated from those without CTCs by chronology horizons. Our aim here is to investigate Li's time machine within the context of modified gravity theories.\\ 

Observational data from type Ia supernovae \cite{Riess, Per}, X-ray experiments \cite{Allen}, large-scale structure \cite{Aba}, and cosmic microwave background fluctuations \cite{Benn} have demonstrated that the Universe is expanding at an accelerated rate. General relativity cannot account for this observational phenomenon unless exotic components are included in the matter content. Einstein's theory faces another challenge: it is a classical theory, and there is no consistent and complete theory of quantum gravity. These two facts, the late-time acceleration of the universe and the lack of a quantum theory of gravity, motivate the research community to explore alternative theories of gravity. Several modified gravitational theories have emerged as significant alternatives to GR, aiming to explain the expansion of the universe and has garnered considerable attention within the scientific community. Notable among these modified gravitational theories are: $f(\mathcal{R})$-gravity, $f(\mathcal{G})$-gravity, where $\mathcal{G}$ is the Gauss-Bonnet invariant, $f(\mathcal{R}, \mathcal{G})$ gravity, $f(\mathcal{R}, \mathcal{T})$ gravity, where $\mathcal{T}$ is the trace of the stress-energy tensor, $f(\mathcal{G}, \mathcal{T})$-gravity, $f(\mathcal{T})$-gravity, $f(Q)$-gravity, where $Q$ is the non-metricity scalar, $f(\mathcal{R}, \mathcal{G}, \mathcal{T})$-gravity, $f(Q, \mathcal{T})$-gravity, and $f(\mathcal{R}, \mathcal{L}_{m})$-gravity, where $\mathcal{L}_m$ is the matter Lagrangian. For a review of modified gravity and the motivation of current study, see references \cite{ref7, ref11, ref6, ref3}. The modified theory explored here is Ricci-inverse gravity.\\

The Ricci-inverse gravity is a modified theory of gravity that incorporates terms involving the anti-curvature tensor into the gravitational action \cite{ref1}.  The anti-curvature tensor and its scalar are defined as $A^{\mu\sigma}\,R_{\nu\sigma}=\delta^\mu_\nu$ and $A=g_{\mu\nu}\,A^{\mu\nu}$, respectively. It is important to note that the anti-curvature scalar is different from the inverse of the Ricci scalar, i.e., $A\neq \mathcal{R}^{-1}$. In a subsequent study documented in Ref. \cite{ref2}, the authors studied into two distinct classes of Ricci-inverse models: Class {\bf I} and Class {\bf II}. The former, denoted as Class {\bf I} models, takes the form $f(\mathcal{R}, A)$, where $f$ is a function of both the Ricci scalar ($\mathcal{R}$) and the anti-curvature scalar ($A$). On the other hand, Class {\bf II} models assume the form $f(\mathcal{R}, A^{\mu\nu}\,A_{\mu\nu})$, where $f$ is a function of the Ricci scalar and the square of the anti-curvature tensor. A notable finding of their investigation is the inadequacy of these two classes of models in explaining the late-time evolution of the Universe. Specifically, the transition from a matter-dominated epoch to an accelerated expansion epoch cannot be accounted for within the frameworks of Class {\bf I} and Class {\bf II} models. In a comprehensive examination presented in Ref. \cite{ref3}, the authors provided an in-depth discussion on the significance of modified gravity theories.\\

Our motivation of current paper is as follows: causality violation has been a significant concern in general relativity since 1949 when Gödel introduced his rotating cosmological model \cite{Godel}. Moreover, there is no rigorous proof of the Hawking's Chronology Protection Conjecture has yet been known in the literature. Taking advantage of non-proofing Conjecture, several researchers have explored causality violating models, presenting solutions with various matter contents, some of which satisfy the energy conditions as stated earlier. Interest in time-machine models peaked when Morris and his collaborators constructed a traversable wormhole \cite{MT1,MT2}, another theoretical possibility for time-machines. All these discussions were primarily conducted within the framework of general relativity. However, due to other significant issues arising within general relativity, physicists have proposed modified gravity theories and analyzed various models. The latest modified gravity theory known to us is the Ricci-inverse gravity theory, an extension of $f(\mathcal{R})$-gravity theory that incorporates a geometric entity called anti-curvature tensor into the Einstein-Hilbert action. We aim to investigate such causality violating space-times constructed in general relativity within the context of Ricci-inverse gravity theory and determine whether causality violation persists. This can be confirmed if solutions to the modified field equations are found. If such solutions exist, it would imply that time-machines are theoretically still possible within Ricci-inverse gravity.\\

Thus, our primary objective of this research is to explore a vacuum time-machine space-time formulated in general relativity, as originally proposed by Li \cite{LX} (see, also \cite{LX2}), within the context of Ricci-inverse gravity. We aim to establish that Li time-machine space-time constitutes a valid vacuum solution within the framework of Ricci-inverse gravity, covering Class-\textbf{I} to Class-\textbf{III } models {\it i.e.}, (i) Class-{I}: $f(\mathcal{R}, \mathcal{A})=(\mathcal{R}+{ \kappa\,\mathcal{R}^2}+\beta\,\mathcal{A})$ model, (ii) Class-{\bf II}: $f(\mathcal{R}, A^{\mu\nu}\,A_{\mu\nu})=(\mathcal{R}+{ \kappa\,\mathcal{R}^2}+\gamma\,A^{\mu\nu}\,A_{\mu\nu})$ model, and (iii) Class-{\bf III}: $f(\mathcal{R}, \mathcal{A}, A^{\mu\nu}\,A_{\mu\nu})=(\mathcal{R}+{\kappa\,R^2}+\beta\,\mathcal{A}+\delta\,\mathcal{A}^2+\gamma\,A^{\mu\nu}\,A_{\mu\nu})$ model, thereby allowing for the existence of closed time-like curves. In fact, it is shown that modified field equations are solvable in all three classes of this gravity theory, and thus, permits the formation of closed time-like curves analogous to general relativity. Therefore, theoretically, time-machine construction is possible within the framework of Ricci-inverse gravity. Moreover, one can see that for $\kappa \neq 0$ and other coupling constants to be zeros, $\beta=0=\gamma=\delta$, one will have modified $f(\mathcal{R})$-gravity theory. We show that the results in GR and $f(\mathcal{R})$-gravity with the function $f(\mathcal{R})=(\mathcal{R}+\kappa\,\mathcal{R}^2)$ are the same. But there are higher order scalar curvature effects $(\mathcal{R}^n, n>2)$ on the results of this $f(\mathcal{R})$-gravity which also we discuss in the current work. It is worthwhile mentioning that this Ricci-inverse gravity theory has gained significant attention in recent studies, as reported in Refs. \cite{epjc,CJPHY,PLB,EPJC2,CJPHY2,NA, EPJC, EPJP, NPB, AOP}.\\

The paper is structured as follows: In Section \ref{sec:2}, we explore a time-machine space-time that allows for closed time-like curves within the framework of general theory of relativity. In Section \ref{sec:3}, we conduct a detailed analysis of this space-time within the context of Ricci-inverse gravity theory, investigating models of Class I, Class II, and Class III. In Section \ref{sec:4}, the geodesic equations are derived and analyzed. In Section \ref{sec:5}, the same time-machine model is examined within the framework of $f(R)$-gravity theory. Finally, in Section \ref{sec:6}, we present some concluding remarks. Throughout the paper, geometric units are used, where $\hbar = c = 8\,\pi\,G = 1$.

\section{Time-machine constructed in GTR }\label{sec:2}

In this part, we study a time-machine model formulated in general theory of relativity (GTR)  by Li \cite{LX}. Therefore, we begin this section by introducing this time-machine space-time by the following line-element in the coordinates ($t, \phi, y, z$) given by \cite{LX}
\begin{equation}
ds^2=-(dt-t\,d\phi)^2+\frac{1}{\alpha^2}\,d\phi^2+(dy-y\,d\phi)^2+(dz-z\,d\phi)^2,\label{Li}
\end{equation}
where different coordinates are in the ranges $-\infty < t < +\infty$, $\phi \sim \phi+2\,n\,\pi$ ($n=\pm\,1,\pm\,2,....$), $-\infty < y < +\infty$, and $-\infty < z < +\infty$. 

The above metric (\ref{Li}) can be expressed in the following form
\begin{equation}
ds^2=g_{tt}\,dt^2+g_{\phi\phi}\,d\phi^2+2\,g_{t\phi}\,dt\,d\phi+g_{yy}\,dy^2+2\,g_{y\phi}\,dy\,d\phi+g_{zz}\,dz^2+2\,g_{z\phi}\,dz\,d\phi,\label{1}
\end{equation}
where the covariant metric tensor $g_{\mu\nu}$ $(\mu,\nu=0,1,2,3)$ is given by 
\begin{eqnarray}
g_{\mu\nu}=\begin{pmatrix}
    -1 & t & 0 & 0\\
    t  & \left(\frac{1}{\alpha^2}+y^2+z^2-t^2\right) & -y & -z\\
    0  & -y & 1 & 0\\
    0  & -z & 0 & 1
    \end{pmatrix}\,.
    \label{2}
\end{eqnarray}
The contravariant metric tensor $g^{\mu\nu}$ is given by 
\begin{eqnarray}
g^{\mu\nu}=\begin{pmatrix}
    -1+\alpha^2\,t^2  & \alpha^2\,t  & \alpha^2\,t\,y & \alpha^2\,t\,z\\
    \alpha^2\,t  & \alpha^2  & \alpha^2\,y  & \alpha^2\,z\\
    \alpha^2\,t\,y  & \alpha^2\,y & 1+\alpha^2\,y^2 & \alpha^2\,y\,z\\
    \alpha^2\,t\,z  & \alpha^2\,z  & \alpha^2\,y\,z & 1+\alpha^2\,z^2
\end{pmatrix}\,.
\label{3}
\end{eqnarray}

The covariant Ricci tensor $R_{\mu\nu}$ for metric (\ref{1}) will be 
\begin{eqnarray}
    R_{\mu\nu}=-3\,\alpha^2\,\begin{pmatrix}
    -1 & t & 0 & 0\\
    t & \big(\frac{1}{\alpha^2}+y^2+z^2-t^2\big) & -y & -z\\
    0 & -y & 1 & 0\\
    0 & -z & 0 & 1
\end{pmatrix}.
\label{4}
\end{eqnarray}
It's contravariant form $R^{\mu\nu}$ will be
\begin{eqnarray}
    R^{\mu\nu}=-3\,\alpha^2\,\begin{pmatrix}
    -1+\alpha^2\,t^2 & t & \alpha^2\,t\,y & \alpha^2\,t\,z\\
    \alpha^2\,t & \alpha^2 & \alpha^2\,y & \alpha^2\,z\\
    \alpha^2\,t\,y & \alpha^2\,y & (1+\alpha^2\,y^2) & \alpha^2\,y\,z\\
    \alpha^2\,t\,z & \alpha^2\,z & \alpha^2\,y\,z & (1+\alpha^2\,z^2)
\end{pmatrix}.
\label{5}
\end{eqnarray}
The Ricci scalar is given by
\begin{equation}
 \mathcal{R}=g_{\mu\nu}\,R^{\mu\nu}=-12\,\alpha^2.\label{6}
\end{equation}

With these ingredients, we find that the Ricci tensor $R_{\mu\nu}$ can be expressed as follows:
\begin{equation}
    R^{\mu\nu}=(-3\,\alpha^2)\,g^{\mu\nu}\label{7}
\end{equation} 
which can be written as
\begin{equation}
    R^{\mu\nu}=\Lambda\,g^{\mu\nu}\quad,\quad \mathcal{R}=4\,\Lambda\label{7a}
\end{equation}
with
\begin{equation}
    \Lambda=-3\,\alpha^2<0\,.\label{7b}
\end{equation}

Therefore, the space-time described by the line-element (\ref{Li}) fulfills the vacuum field equations with a negative cosmological constant in general relativity, thus, representing an Anti-de Sitter (AdS) model.\\

Now, we discuss the causality violation issue on space-time (\ref{Li}). For that considering a curve defined by $y=y_0$ and $z=z_0$, where $y_0$ and $z_0$ are arbitrary constants. Therefore, we finds a two-dimensional Misner-like space given by
\begin{equation}
    ds^2=-dt^2+\Big(\frac{1}{\alpha^2}+y^2_{0}+z^2_{0}-t^2\Big)\,d\phi^2+2\,t\,dt\,d\phi.\label{E3}
\end{equation}

The closed curve $\Big\{\phi \sim \phi+2\,n\,\pi,\,\, y=y_0,\,\, z=z_0\Big\}$ ($n=\pm\,1,\pm\,2,....$) is time-like in the region $t^2>\frac{1}{\alpha^2}+y^2_{0}+z^2_{0}$,\, space-like in the region $t^2 < \frac{1}{\alpha^2}+y^2_{0}+z^2_{0}$,\, and light-like at $t^2=\frac{1}{\alpha^2}+y^2_{0}+z^2_{0}$. Hence, vacuum space-time (\ref{Li}) admits closed time-like curves in the time-like region defined by $t^2> \frac{1}{\alpha^2}+y^2_{0}+z^2_{0}$, and therefore, violates the causality condition. We conclude that space-time (\ref{Li}) acts as time-machine model theoretically.\\

One can write the above line-element (\ref{Li}) in terms of the cosmological constant $\Lambda$ as follows:
\begin{equation}
    ds^2=-(dt-t\,d\phi)^2+\frac{3}{(-\Lambda)}\,d\phi^2+(dy-y\,d\phi)^2+(dz-z\,d\phi)^2.\label{9}
\end{equation}
For this space-time, the contravariant metric tensor $g^{\mu\nu}$ is given by 
\begin{eqnarray}
g^{\mu\nu}=-\frac{\Lambda}{3}\,\begin{pmatrix}
    t^2+\frac{3}{\Lambda}  & t  & t\,y & t\,z\\
    t  & 1  & y  & z\\
    t\,y  & y & y^2-\frac{3}{\Lambda} & y\,z\\
    t\,z  & z  & y\,z & z^2-\frac{3}{\Lambda}
\end{pmatrix}.\label{matrix}
\end{eqnarray}

Below, we will discuss this space-time (\ref{9}) within the context of Ricci-inverse modified gravity and show that this space-time admits closed time-like curves in both theories, {\it i.e.}, general theory of relativity and Ricci-inverse gravity, thereby functioning as time-machine model theoretically.  

\section{ Time-machines construct within RI-gravity theory }\label{sec:3}

In this section, we intend to investigate the space-time described by (\ref{9}) within the framework of modified gravity theories, specifically Ricci-inverse gravity \cite{ref1,ref2}. It's important to note that in Ricci-inverse gravity model, the determinant of the Ricci tensor $R_{\mu\nu}$ associated with any metric tensor must be non-zero. In our case, we observe that the determinant of the Ricci tensor $R_{\mu\nu}$ for the line-element (\ref{9}) is non-zero, as given by
\begin{eqnarray}
    \mbox{det}\,(R_{\mu\nu})=3\,\Lambda^3.\label{10}
\end{eqnarray}
Therefore, there exist an anti-curvature tensor $A^{\mu\nu}$ defined by
\begin{eqnarray}
    A^{\mu\nu}=R^{-1}_{\mu\nu}=\frac{\mbox{adj}\,(R_{\mu\nu})}{\mbox{det}\,(R_{\mu\nu})}.\label{11}
\end{eqnarray}
For space-time (\ref{9}), we obtain this anti-curvature tensor $A^{\mu\nu}$ and its covariant form $A_{\mu\nu}$ given by
\begin{eqnarray}
    &&A^{\mu\nu}=-\frac{1}{3}\begin{pmatrix}
        \frac{3}{\Lambda}+t^2 & t  & t\,y & t\,z\\
        t & 1 & y & z \\
        t\,y & y & -\frac{3}{\Lambda}+y^2 & y\,z\\
        t\,z & z & y\,z & -\frac{3}{\Lambda}+z^2
    \end{pmatrix},\nonumber\\
    &&A_{\mu\nu}=\frac{1}{(-\Lambda)}\begin{pmatrix}
    1 & -t & 0 & 0\\
    -t & t^2+\frac{3}{\Lambda}-y^2-z^2 & y & z\\
    0 & y & -1 & 0\\
    0 & z & 0 & -1
    \end{pmatrix}.\label{12}
\end{eqnarray}
The anti-curvature scalar is given by
\begin{equation}
    \mathcal{A}=g_{\mu\nu}\,A^{\mu\nu}=\frac{4}{\Lambda}\,.\label{13}
\end{equation}

We will now explore how this time-machine space-time fits within the different classes of Ricci-inverse theory.

\subsection{ Time-machine construct in RI-gravity: Class-{\bf I} Models } \label{III1}

We introduce anti-curvature tensor $A^{\mu\nu}$ into the Lagrangian of the system in general relativity theory. Therefore, the action that describes the Ricci-inverse gravity is given by \cite{ref1,ref2}
\begin{eqnarray}
    S= \int d^4x\, \sqrt{-g}\left[{ f(\mathcal{R},\mathcal{A})}-2\,\Lambda_m+{\cal L}_m\right],\label{14}
\end{eqnarray}
where $\Lambda_m$ is the cosmological constant in this new theory, ${\cal L}_m$ is the Lagrangian of the matter content, and other symbols have their usual meanings. This action can be written as
\begin{eqnarray}
    S= \int d^4x \sqrt{-g}\left[{ f(g_{\mu\nu}\,R^{\mu\nu}, g_{\mu\nu}\,A^{\mu\nu}})-2\,\Lambda_m+{\cal L}_m\right].\label{14b}
\end{eqnarray}

Now, varying the action (\ref{14b}) with respect to the metric tensor { $g_{\mu\nu}$, the modified} field equations that describe this Ricci-inverse gravity theory is given by
\begin{eqnarray}
    f_{\mathcal{R}}\,R^{\mu\nu}-\frac{f}{2}\,g^{\mu\nu}+\Lambda_m\,g^{\mu\nu}+N^{\mu\nu}=\mathcal{T}^{\mu\nu},\label{15}
\end{eqnarray}
with $f_{\mathcal{R}}=\partial f/\partial \mathcal{R}$,\, $\mathcal{T}^{\mu\nu}$ being the standard energy-momentum tensor and $N^{\mu \nu}$ which is symmetric is defined as
\begin{eqnarray}
     N^{\mu\nu}&=&\frac{1}{2}\,\Big[2\,g^{\rho\mu}\nabla_{\iota}\,\nabla_{\rho} ({ f_{\mathcal{A}}}\,A^{\iota}_{\sigma}\,A^{\nu\sigma})-\nabla^2({ f_{\mathcal{A}}}\,A^{\mu}_{\iota}\,A^{\nu\iota})-g^{\mu\nu}\,\nabla_{\rho}\,\nabla_{\iota}({ f_{\mathcal{A}}}\,A^{\rho}_{\sigma}\,A^{\iota\sigma})\Big]\nonumber\\
     &+&{ g^{\mu\nu}\,\nabla^{\sigma}\nabla_{\sigma}\,f_{\mathcal{R}}-\nabla^{\mu}\nabla^{\nu}\,f_{\mathcal{R}}}-{ f_{\mathcal{A}}}\,A^{\mu\nu}.\label{16}
\end{eqnarray} 
Here, we have used $A^{\tau}_{\sigma}\,A^{\nu\sigma}=A^{\tau\lambda}\,g_{\lambda\sigma}\,A^{\nu\sigma}=A^{\tau\lambda}\,A^{\nu}_{\lambda}$ and $f_{\mathcal{A}}=\partial f/\partial \mathcal{A}$.

In this Class-{\bf I} model, we choose the following function
\begin{equation}
    f(\mathcal{R},\mathcal{A})=(\mathcal{R} +\kappa\,\mathcal{R}^2+\beta\,\mathcal{A}),\label{function1}
\end{equation}
where $\kappa$ and $\beta$ are the arbitrary coupling constants. Noted that for $\beta=0$, one will have modified $f(\mathcal{R})$-gravity theory.

Using the above function, we finds the following
\begin{eqnarray}
    f_{\mathcal{R}}&=&1+2\,\kappa\,\mathcal{R},\nonumber\\
    f_{\mathcal{A}}&=&\beta.\label{function2}
\end{eqnarray}
Using the Ricci tensor and the Ricci scalar given in equation (\ref{7a}), we can write the modified field equations (\ref{15}) as follows
\begin{eqnarray}(1+2\,\kappa\,\mathcal{R})\,R^{\mu\nu}-\frac{1}{2}\,(\mathcal{R}+\kappa\,\mathcal{R}^2+\beta\,\mathcal{A})\,g^{\mu\nu}+\Lambda_m\,g^{\mu\nu}+N^{\mu \nu}=\mathcal{T}^{\mu\nu},\label{17}
\end{eqnarray}
where $\nabla^{\sigma}\nabla_{\sigma}\,(1+2\,\kappa\,\mathcal{R})=0=\nabla^{\mu}\nabla^{\nu}\,(1+2\,\kappa\,\mathcal{R})$ since the Ricci scalar $\mathcal{R}$ is a constant and 
\begin{eqnarray}
     N^{\mu \nu}=-\beta\,A^{\mu\nu}+\frac{\beta}{2}\,\Big[2\,g^{\rho\mu}\nabla_{\iota}\,\nabla_{\rho} (A^{\iota}_{\sigma}\,A^{\nu\sigma})-\nabla^2(A^{\mu}_{\iota}\,A^{\nu\iota})-g^{\mu\nu}\,\nabla_{\rho}\,\nabla_{\iota}(A^{\rho}_{\sigma}\,A^{\iota\sigma})\Big].\label{function3}
\end{eqnarray}

Now, we need to calculate $N^{\mu \nu}$ using the anti-curvature tensor given in Eq. (\ref{12}), the anti-curvature scalar given in (\ref{13}), and the metric tensor (\ref{matrix}). This tensor $N^{\mu \nu}$ and its covariant form $N_{\mu \nu}$ are given by:
\begin{eqnarray}
    &&N^{\mu \nu}=\beta\,\begin{pmatrix}
        t^2+\frac{3}{\Lambda} & t & t\,y & t\,z\\
        t & 1 & y & z\\
        t\,y & y & -\frac{3}{\Lambda}+y^2 & y\,z\\
        t\,z & z & y\,z & -\frac{3}{\Lambda}+z^2
    \end{pmatrix}\nonumber\\
    &&N_{\mu \nu}=\frac{3\,\beta}{(-\Lambda)}\,\begin{pmatrix}
        -1 & t & 0 & 0\\
        t & -\frac{3}{\Lambda}+y^2+z^2-t^2 & -y & -z\\
        0 & -y & 1 & 0\\
        0 & -z & 0 & 1
    \end{pmatrix}.\label{18}
\end{eqnarray}

Substituting $g^{\mu\nu}$ from Eq. (\ref{matrix}) and $N^{\mu \nu}$ from Eq. (\ref{18}) into the modified field equations (\ref{17}), we obtain
\begin{eqnarray}
    \mathcal{T}^{\mu\nu}&=&\frac{\Lambda}{3}\left(\frac{3\,\beta}{\Lambda}+\Lambda-\Lambda_m\right)\begin{pmatrix}
        t^2+\frac{3}{\Lambda} & t & t\,y & t\,z \\
        t & 1 & y & z \\
        t\,y & y & y^2-\frac{3}{\Lambda} & y\,z\\
        t\,z & z & y\,z & z^2-\frac{3}{\Lambda}
    \end{pmatrix}\nonumber\\
    &=&{ -\left(\frac{3\,\beta}{\Lambda}+\Lambda-\Lambda_m\right)g^{\mu\nu}}.\label{19}
\end{eqnarray}

Considering vacuum as matter content whose energy-momentum tensor is $\mathcal{T}^{\mu\nu}=0$ in the modified theory of gravity. Thereby, solving the modified field equations (\ref{17}) in vacuum, we obtain the following relation 
\begin{eqnarray}
    \Big(\frac{3\,\beta}{\Lambda}+\Lambda-\Lambda_m\Big)=0,\quad \Lambda \neq 0\,.\label{20}
\end{eqnarray}
Simplification of the above equation results an effective cosmological constant given by
\begin{equation}
    \Lambda_m=\Lambda+\frac{3\,\beta}{\Lambda}\,.\label{21}
\end{equation}

It is evident that when the coupling constant $\beta$ approaches zero ($\beta \to 0$), an effective cosmological constant simplifies to the original one, $\Lambda_m \to \Lambda$. Thus, we see that the line-element (\ref{9}) serves as a valid solution in Ricci-inverse modified gravity theory in Class-{\bf I} models. Since $\Lambda$ in GR is negative, thus, an effective cosmological constant $\Lambda_m$ is either positive or negative provided the following condition must obey:

\begin{itemize}
    \item $\Lambda <0$,\quad $\beta>0$, \quad $\Lambda_m<0$.
    \item $\Lambda <0$,\quad $\beta<0$, \quad $\Lambda_m<0$ \quad provided\quad $\beta<-\Lambda^2/3$.
    \item $\Lambda <0$,\quad $\beta<0$, \quad $\Lambda_m>0$ \quad provided\quad $\beta>-\Lambda^2/3$. 
\end{itemize}

Therefore, we conclude that the space-time described by the line-element (\ref{9}) is a valid solution in Ricci-inverse gravity theory of Class-{\bf I} models with an effective cosmological constant given in Eq. (\ref{21}).

\subsection{ Time-machines construct in RI-gravity: Class-{\bf II} Models }\label{III2}

In this part, we discuss Class-{\bf II} models of Ricci-inverse gravity theory using the space-time (\ref{9}). The action that describes this $\mathcal{RI}$ gravitational theory of Class-II models is given as
\begin{eqnarray}
    S&=& \int d^4x\, \sqrt{-g}\left[f(\mathcal{R}, A^{\mu\nu}\,A_{\mu\nu})-2\,\Lambda+{\cal L}_m\right],\quad \mathcal{R}=g_{\mu\nu}\,R^{\mu\nu}\,.\label{A1}
\end{eqnarray}

To obtain the modified field equations, we perform a variation with respect to the metric tensor $g_{\mu\nu}$, leading to the following equations
\begin{eqnarray}
   \nonumber &&  f_{\mathcal{R}}\,R^{\mu \nu} - \frac{1}{2}\,f\,g^{\mu \nu}+\Lambda\, g^{\mu\nu} -  2\,f_{A^2}\,A^{\rho \nu}\,A^{\mu}_{\rho}     
 -  \nabla ^{\mu}\, \nabla ^{\nu}\,f_{\mathcal{R}} +  g^{\mu \nu}\, \nabla ^{\lambda}\,\nabla _{\lambda}\,f_{\mathcal{R}}  \nonumber\\
 &+& g^{\rho \nu}\, \nabla _{\alpha}\,  \nabla _{\rho}\,( f_{A^2}\,A_{\sigma \kappa}\, A^{\sigma \alpha}\,A^{\kappa \mu})    
 - \nabla ^2\,(f_{A^2}\,A_{\sigma \kappa}\, A^{\sigma \mu }\,A^{\kappa \nu}) - g^{\mu \nu}\, \nabla _{\alpha}\, \nabla _{\rho}\,(f_{A^2}\,A_{\sigma \kappa}\, A^{\sigma \alpha}\,A^{\kappa \rho})    \nonumber\\ 
     &+& 2\,g^{\rho \nu}\, \nabla _{\rho} \nabla _{\alpha}( f_{A^2}\,A_{\sigma \kappa}\, A^{\sigma \mu }\,A^{\kappa \alpha} )    -g^{\rho \nu}\, \nabla _{\alpha} \nabla _{\rho}\,(f_{A^2}\,A_{\sigma \kappa}\, A^{\sigma \mu}\,A^{\kappa \alpha})     =\mathcal{T}^{\mu\nu} \, ,\label{A2}
\end{eqnarray}
with $\mathcal{T}^{\mu\nu}$ being the energy-momentum tensor and we have considered 
\begin{equation}\label{A3}
    f=f(\mathcal{R}, A^{\mu\nu}\,A_{\mu\nu}),\quad f_{\mathcal{R}}=\partial f/\partial \mathcal{R},\quad f_{A^2}=\partial f/\partial (A^{\mu\nu}\,A_{\mu\nu})\,.
\end{equation}

In order to express the field equations (\ref{A2}) more succinctly, let's define the following tensors
\begin{eqnarray}
Y^{\mu \nu}\equiv 2\,f_{A^2}\,A^{\rho \nu}\,A^{\mu}_{\rho}\,. \label{A4}
\end{eqnarray}
And 
\begin{eqnarray}
    \nonumber U^{\mu \nu}&\equiv& - \nabla ^{\mu} \nabla ^{\nu}f_{\mathcal{R}} +  g^{\mu \nu}\, \nabla ^{\lambda}\,\nabla _{\lambda}f_{\mathcal{R}}        + g^{\rho \nu}\, \nabla _{\alpha}  \nabla _{\rho}( f_{A^2}\,A_{\sigma \kappa}\, A^{\sigma \alpha}\,A^{\kappa \mu})    \\
   \nonumber &-& \nabla ^2(f_{A^2}\,A_{\sigma \kappa}\, A^{\sigma \mu }\,A^{\kappa \nu}) - g^{\mu \nu}\, \nabla _{\alpha} \nabla _{\rho}(f_{A^2}\,A_{\sigma \kappa}\, A^{\sigma \alpha}\,A^{\kappa \rho})      \\ 
     &+& 2\,g^{\rho \nu}\, \nabla _{\rho} \nabla _{\alpha}( f_{A^2}\,A_{\sigma \kappa}\, A^{\sigma \mu }\,A^{\kappa \alpha})-g^{\rho \nu}\, \nabla _{\alpha} \nabla _{\rho}(f_{A^2}\,A_{\sigma \kappa}\,A^{\sigma \mu}\,A^{\kappa \alpha}) \, .\label{A5}
\end{eqnarray}

Then the field equation (\ref{A2}) becomes
\begin{eqnarray}\label{A6}
  f_{\mathcal{R}}\, R^{\mu \nu} - \frac{1}{2}\,f\,g^{\mu \nu}+\Lambda\, g^{\mu\nu} -  Y^{\mu \nu}  +U^{\mu \nu} =\mathcal{T}^{\mu \nu}\,.
\end{eqnarray}

Let's consider the function $f$ in this Class-{\bf II} to be the following form:
\begin{equation} \label{A7}
    f(\mathcal{R}, A^{\mu \nu}\,A_{\mu \nu}) =\mathcal{R}+{ \kappa\,\mathcal{R}^2 }+\gamma\, A^{\mu \nu}\,A_{\mu \nu} \, 
\end{equation}
with $\gamma$ and { $\kappa$} being arbitrary constants. This leads to
\begin{eqnarray}\label{A8}
    \nonumber f_{\mathcal{R}}= 1+{ 2\,\kappa\,\mathcal{R}},\quad\quad f_{A^2}= \gamma.
\end{eqnarray}

Noted that one can consider higher order terms of both the Ricci scalar and the quadratic invariant. In this analysis, we have considered the above function form (\ref{A7}) and let's see whether we will get a solution in this Class-{\bf II} of $\mathcal{RI}$-gravity.\\

Substituting function (\ref{A7}) into the Eq. (\ref{A6}) results the following field equations given by
\begin{equation}\label{A9}
    { (1+2\,\kappa\,\mathcal{R})}\,R^{\mu\nu}-\frac{1}{2}\,(\mathcal{R}+{ \kappa\,\mathcal{R}^2})\,g^{\mu\nu}-\frac{\gamma}{2}\,A^{\mu\nu}\,A_{\mu\nu}\,g^{\mu\nu}-2\,\gamma\,A^{\rho\nu}\,A^{\mu}_{\rho}+\tilde{U}^{\mu\nu}+\Lambda_{m}\,g^{\mu\nu}=\mathcal{T}^{\mu\nu},
\end{equation}
where
\begin{eqnarray}\label{A10}
    \tilde{U}^{\mu\nu}&=&\gamma\,\Big[g^{\rho \nu} \nabla _{\alpha}  \nabla _{\rho}(A_{\sigma \kappa} A^{\sigma \alpha}A^{\kappa \mu})-\nabla ^2(A_{\sigma \kappa} A^{\sigma \mu }A^{\kappa \nu})-g^{\mu \nu} \nabla _{\alpha} \nabla _{\rho}(A_{\sigma \kappa} A^{\sigma \alpha}A^{\kappa \rho}) \nonumber\\
   &+&2\,g^{\rho \nu} \nabla _{\rho} \nabla _{\alpha}(A_{\sigma \kappa} A^{\sigma \mu }A^{\kappa \alpha})-g^{\rho \nu} \nabla _{\alpha} \nabla _{\rho}(A_{\sigma \kappa} A^{\sigma \mu}A^{\kappa \alpha})\Big]\,.
\end{eqnarray}

Substituting the Ricci tensor $R_{\mu\nu}$ and the Ricci scalar $\mathcal{R}$ given in Eq. (\ref{7a}), and the anti-curvature tensor $A^{\mu\nu}$ given in Eq. (\ref{12}) into the modified fields equations (\ref{A9}), we obtain
\begin{eqnarray}\label{A11}
    \mathcal{T}^{\mu\nu}&=&\frac{\Lambda}{3}\,\left(\Lambda+\frac{4\,\gamma}{\Lambda^2}-\Lambda_m\right)\,\begin{pmatrix}
        t^2+\frac{3}{\Lambda} & t & t\,y & t\,z\\
        t & 1 & y & z\\
        t\,y & y & y^2-\frac{3}{\Lambda} & y\,z\\
        t\,z & z & y\,z & z^2-\frac{3}{\Lambda}
    \end{pmatrix}\nonumber\\
    &=&{ -\left(\Lambda+\frac{4\,\gamma}{\Lambda^2}-\Lambda_m\right)g^{\mu\nu}}.
\end{eqnarray}

Solving the modified field equations (\ref{A11}) for vacuum as matter-energy content, where $\mathcal{T}^{\mu\nu}=0$, we obtain
\begin{equation}
    \left(\Lambda-\Lambda_m+\frac{4\,\gamma}{\Lambda^2}\right)=0\Rightarrow \Lambda_m=\Lambda+\frac{4\,\gamma}{\Lambda^2}\,.\label{A12}
\end{equation}

From the above analysis, one can see that the coupling constant approaches zero, $\gamma=0$, an effective cosmological constant simplifies to the general relativity one, $\Lambda_m \to \Lambda$. Thus, the line-element (\ref{9}) serves as a valid solution in Ricci-inverse gravity theory of Class-{\bf II} models with the function $f(\mathcal{R}, A^{\mu \nu}\,A_{\mu \nu}) = (\mathcal{R}+\gamma\, A^{\mu \nu}\,A_{\mu \nu})$. Since $\Lambda$ in GR is negative, thus, an effective cosmological constant $\Lambda_m$ is either positive or negative provided the following condition must obey:

\begin{itemize}
    \item $\Lambda <0$,\quad $\gamma>0$, \quad $\Lambda_m<0$\quad provided\quad $\gamma<-\Lambda^3/4$.
    \item $\Lambda <0$,\quad $\gamma>0$, \quad $\Lambda_m>0$ \quad provided\quad $\gamma>-\Lambda^3/4$.
    \item $\Lambda <0$,\quad $\gamma<0$, \quad $\Lambda_m<0$.
\end{itemize}

\subsection{Time-machines construct in RI-gravity: Class-{\bf III} Models }\label{III3}

In this part, we explore Class-{\bf III} models of Ricci-inverse gravity, where the function $f$ in the gravitational action takes the form $f(\mathcal{R}, \mathcal{A}, A^{\mu \nu}\,A_{\mu \nu})$. 

In this way, the action describing this theory in Class-{\bf III} models of $\mathcal{RI}$-gravity is given as
\begin{equation}\label{B1}
    S = \int \mathrm{d}x^4 \sqrt{-g}\,[ f(\mathcal{R},\mathcal{A},A^{\mu \nu}A_{\mu \nu})-2\,\Lambda ]+ S_M\,.
\end{equation}

By varying the action (\ref{B1}) with respect to the metric tensor $g_{\mu\nu}$, the modified field equations is given by  
\begin{equation}\label{B2}
    -\frac{1}{2}\, f\, g^{\mu \nu} + f_{\mathcal{R}} \, R^{\mu \nu}-f_{\mathcal{A}} \, A^{\mu \nu} - 2\,f_{A^2}\, A^{\rho \nu }A^{\mu}_{\rho} +P^{\mu \nu}+ M^{\mu \nu} + U^{\mu \nu} + \Lambda g^{\mu \nu} =\mathcal{T}^{\mu \nu} \, ,
\end{equation}
where 
\begin{eqnarray}
    P^{\mu \nu} &=& g^{\mu \nu} \nabla ^2 \, f_{\mathcal{R}}-\nabla ^{\mu} \nabla^{\nu}\, f_{\mathcal{R}} \, ,\label{B3}\\ 
    M^{\mu \nu} &=& g^{\rho \mu}\nabla _{\alpha } \nabla _{\rho } (f_{\mathcal{A}}\, A_{\sigma}^{ \alpha}\, A^{\nu \sigma}) - \frac{1}{2}\, \nabla ^2 (f_{\mathcal{A}}\,A^{\mu}_{\sigma}\, A^{\nu \sigma}) - \frac{1}{2}\, g^{\mu \nu}\, \nabla _{\alpha} \nabla _{ \rho} ( f_{\mathcal{A}}\,A_{\sigma}^{ \alpha}\, A^{\rho \sigma})\, ,\label{B4}\\
 \nonumber  U^{\mu \nu} &=& g^{\rho \nu}\,\nabla _{\alpha} \nabla _{\rho}(f_{A^2}\,A_{\sigma \kappa}\,A^{\sigma \alpha}\,A^{\mu \kappa})-\nabla ^{2}(f_{A^2}\,A_{\sigma \kappa}\,A^{\sigma \mu}\,A^{\nu \kappa})
  -g^{\mu \nu}\,\nabla _{\alpha} \nabla _{\rho}(f_{A^2}\,A_{\sigma \kappa}\,A^{\sigma \alpha}\,A^{\rho \kappa})\nonumber\\
  &+& 2\,g^{\rho \nu}\,\nabla _{\rho} \nabla _{\alpha}(f_{A^2}\,A_{\sigma \kappa}\,A^{\sigma \mu}\,A^{\alpha \kappa})
   - g^{\rho \nu}\,\nabla _{\alpha} \nabla _{\rho}(f_{A^2}\,A_{\sigma \kappa}\,A^{\sigma \mu}\,A^{\alpha \kappa})\, .\label{B5} 
\end{eqnarray}
Here various symbols are defined by
\begin{equation}
    f_{\mathcal{R}}=\partial f/\partial \mathcal{R},\quad f_{\mathcal{A}}=\partial f/\partial \mathcal{A},\quad f_{A^2}=\partial f/\partial (A^{\mu\nu}\,A_{\mu\nu})\,.\label{B6}
\end{equation}

Let's consider the function $f$ in this Class-{\bf III} to be the following form:
\begin{equation} \label{B7}
    f(\mathcal{R},\mathcal{A},A^{\mu \nu}\,A_{\mu \nu}) =\mathcal{R}+{ \kappa\,\mathcal{R}^2}+\beta \, \mathcal{A} + \delta\,\mathcal{A}^2++\gamma\,A^{\mu \nu}\,A_{\mu \nu} \, 
\end{equation}
with $\beta, \delta$ and $\gamma$ being arbitrary constants. Using this function, one can find the following
\begin{eqnarray}\label{B8}
    \nonumber f_{\mathcal{R}}&=& 1+{ 2\,\kappa\,\mathcal{R}} \, , \\
    \nonumber f_{\mathcal{A}}&=& \beta +2\,\delta\,\mathcal{A}\, ,\\
    f_{A^2}&=& \gamma\, .
\end{eqnarray}

Therefore, the modified field equations (\ref{B2}) using (\ref{B7}) reduces to the following form:
\begin{eqnarray}\label{B9}
    &&-\frac{1}{2}\,(\mathcal{R}+{ \kappa\,\mathcal{R}^2}+\beta \, \mathcal{A} + \delta\,\mathcal{A}^2++\gamma\,A^{\mu \nu}\,A_{\mu \nu})\, g^{\mu \nu} +{ (1+2\,\kappa\,\mathcal{R})}\, R^{\mu \nu}-(\beta+2\,\delta\,\mathcal{A})\, A^{\mu \nu}\nonumber\\
    &&- 2\,\gamma\, A^{\rho \nu }\,A^{\mu}_{\rho}+\tilde{M}^{\mu \nu} + \tilde{U}^{\mu \nu} + \Lambda g^{\mu \nu} =\mathcal{T}^{\mu \nu} \, ,
\end{eqnarray}
where $P^{\mu \nu}=0$ and 
\begin{eqnarray}
    \tilde{M}^{\mu \nu} &=& g^{\rho \mu}\,\nabla _{\alpha } \nabla _{\rho }\, [(\beta+2\,\delta\,\mathcal{A})\,A_{\sigma}^{ \alpha}\,A^{\nu \sigma}]-\frac{1}{2}\,\nabla ^2 \,[(\beta+2\,\delta\,\mathcal{A})\,A^{\mu}_{\sigma}\,A^{\nu \sigma}]\nonumber\\
    &-&\frac{1}{2}\,g^{\mu \nu}\,\nabla _{\alpha} \nabla _{ \rho}\, [ (\beta+2\,\delta\,\mathcal{A})\,A_{\sigma}^{ \alpha}A^{\rho \sigma}] ,\label{B10}\\
 \nonumber \tilde{U}^{\mu \nu} &=&\gamma\,\Big[g^{\rho \nu}\,\nabla _{\alpha} \nabla _{\rho}(A_{\sigma \kappa}\,A^{\sigma \alpha}\,A^{\mu \kappa})-\nabla ^{2}(A_{\sigma \kappa}\,A^{\sigma \mu}\,A^{\nu \kappa})
  -g^{\mu \nu}\,\nabla _{\alpha} \nabla _{\rho}(A_{\sigma \kappa}\,A^{\sigma \alpha}\,A^{\rho \kappa})\nonumber\\
  &+& 2\,g^{\rho \nu}\,\nabla _{\rho} \nabla _{\alpha}(A_{\sigma \kappa}\,A^{\sigma \mu}\,A^{\alpha \kappa})
   - g^{\rho \nu}\,\nabla _{\alpha} \nabla _{\rho}(A_{\sigma \kappa}\,A^{\sigma \mu}\,A^{\alpha \kappa})\Big]\,.\label{B11} 
\end{eqnarray}

Substituting the Ricci tensor $R_{\mu\nu}$ and the Ricci scalar $\mathcal{R}$ given in Eq. (\ref{7a}), and the anti-curvature tensor $A^{\mu\nu}$ given in Eq. (\ref{12}) and anti-curvature scalar $\mathcal{A}$ into the modified fields equations (\ref{B9}), we obtain
\begin{eqnarray}\label{B12}
    \mathcal{T}^{\mu\nu}&=&\frac{\Lambda}{3}\,\left(\Lambda+\frac{3\,\beta}{\Lambda}+\frac{4\,\gamma+16\,\delta}{\Lambda^2}-\Lambda_m\right)\,\begin{pmatrix}
        t^2+\frac{3}{\Lambda} & t & t\,y & t\,z\\
        t & 1 & y & z\\
        t\,y & y & y^2-\frac{3}{\Lambda} & y\,z\\
        t\,z & z & y\,z & z^2-\frac{3}{\Lambda}
    \end{pmatrix}\nonumber\\
    &=&{ -\left(\Lambda+\frac{3\,\beta}{\Lambda}+\frac{4\,\gamma+16\,\delta}{\Lambda^2}-\Lambda_m\right) g^{\mu\nu}.}
\end{eqnarray}

Solving the modified field equations (\ref{B11}) for vacuum as matter-energy content, where $\mathcal{T}^{\mu\nu}=0$, using Eq. (\ref{B12}) we obtain
\begin{equation}
    \left(\Lambda+\frac{3\,\beta}{\Lambda}+\frac{4\,\gamma+16\,\delta}{\Lambda^2}-\Lambda_m\right)=0,\quad \Lambda \neq 0\,.\label{B13}
\end{equation}
Simplification of the above relation results
\begin{equation}
    \Lambda_m=\Lambda+\frac{3\,\beta}{\Lambda}+\frac{4\,\gamma}{\Lambda^2}+\frac{16\,\delta}{\Lambda^2}.\label{B14}
\end{equation}

We see that the space-time (\ref{9}) serves as a valid solution in Ricci-inverse gravity theory of Class-{\bf III} models. The cosmological constant gets modifications by the coupling constants $\beta, \gamma$ and $ \delta$. We see that for $\gamma=0=\delta$, we will get the result obtained in Class-{\bf I} models. In addition, for $\beta=0=\delta$, we recover the results of Class-{\bf II} models. an effective cosmological constant $\Lambda_m$ becomes negative provided the following constraints on the coupling constants satisfies:

\begin{itemize}
    \item $\Lambda<0$,\quad $\beta>0$,\quad $\gamma>0$,\quad $\delta>0$,\quad $\Lambda_m<0$\quad\mbox{provided}\quad $\gamma +4\,\delta< \frac{\Lambda^2}{4}\Big(-\Lambda-\frac{3\,\beta}{\Lambda}\Big)$.
    \item $\Lambda<0$,\quad $\beta<0$,\quad $\gamma>0$,\quad $\delta>0$,\quad $\Lambda_m<0$\quad \mbox{provided}\quad $\frac{3\,\beta}{\Lambda}+\frac{4\,\gamma+16\,\delta}{\Lambda^2}<-\Lambda$ .
    \item $\Lambda<0$,\quad $\beta>0$,\quad $\gamma<0$,\quad $\delta>0$,\quad $\Lambda_m<0$\quad \mbox{provided}\quad $\delta <\frac{\Lambda^2}{16}\,\Big(-\Lambda-\frac{3\,\beta}{\Lambda}-\frac{4\,\gamma}{\Lambda^2}\Big)$.
    \item $\Lambda<0$,\quad $\beta <0$,\quad $\gamma<0$,\quad $\delta>0$,\quad $\Lambda_m<0$\quad \mbox{provided}\quad $\frac{16\,\delta}{\Lambda^2}+\frac{3\,\beta}{\Lambda} <\Big(-\Lambda-\frac{4\,\gamma}{\Lambda^2}\Big)$.
    \item $\Lambda<0$,\quad $\beta >0$,\quad $\gamma>0$,\quad $\delta<0$,\quad $\Lambda_m<0$\quad \mbox{provided}\quad $\gamma< \frac{\Lambda^2}{4}\,\Big(-\Lambda-\frac{16\,\delta}{\Lambda^2}-\frac{3\,\beta}{\Lambda} \Big)$.
\end{itemize}

\begin{table}[htb!]
    \centering
\begin{tabular}{|c|c|c|c|c|}
        \hline
        $\mbox{Classes} \to $ & $\mbox{Class}-{\bf I}$ & $\mbox{Class}-{\bf II}$ & $\mbox{Class}$-{\bf III} & $\mbox{GR}$ \cite{LX} \\ [1.0ex] 
        \hline
        $\mbox{Results}$ & $\Lambda+\frac{3\,\beta}{\Lambda}$ & $\Lambda+\frac{4\,\gamma}{\Lambda^2}$ & $\Lambda+\frac{3\,\beta}{\Lambda}+\frac{4\,\gamma+16\,\delta}{\Lambda^2}$ & $\Lambda<0$\\ [2.0ex] 
        \hline
    \end{tabular}
    \caption{The results of Ricci-inverse gravity in Class-{\bf I} to Class-{\bf III} models and GR. }
    \label{table:1}
\end{table}

\begin{small}
\begin{table}[htb!]
    \centering
    \begin{tabular}{|c|c|c|c|c|}
    \hline 
    Refs. & Petrov Classification & Matter content & Energy-density ($\rho$) & RI-gravity \\[1.0ex]
    \hline
    \cite{EPJP} & AdS & vacuum & --& Valid\\[2.0ex]
    \hline
    \cite{epjc, EPJC} & Type-III & pure radiation & depends on $r$ & Valid\\ [2.0ex]
    \hline
    Current & AdS & vacuum & --- & Valid\\[2.0ex]
    \hline
    \end{tabular}
    \caption{Different space-time geometries that are valid solutions in Ricci-inverse gravity theory.}
    \label{table:2}
\end{table}
\end{small}

In Table (\ref{table:1}), we summarized our results obtained in all Classes of Ricci-inverse gravity theory. Table \ref{table:2} compares the results obtained in this work with those of other different space-time geometries within Ricci-inverse gravity theory.

Therefore, space-time (\ref{9}) in the general Class of Ricci-inverse gravity theory with the function $f(\mathcal{R},\mathcal{A},A^{\mu \nu}\,A_{\mu \nu}) =(\mathcal{R}+\kappa\,\mathcal{R}^2+\beta\,\mathcal{A} + \delta\,\mathcal{A}^2++\gamma\,A^{\mu \nu}\,A_{\mu \nu})$ can be described by the following line-element  
\begin{equation}
    ds^2=-(dt-t\,d\phi)^2+\frac{3}{(-\Lambda_m)}\,d\phi^2+(dy-y\,d\phi)^2+(dz-z\,d\phi)^2,\label{special-metric}
\end{equation}
where $\Lambda_m$ is given in Eq. (\ref{B14}). This space-time also represents a vacuum solution in this modified gravity theory.\\

An interesting point we have observed here is that there is no contribution from the term $\mathcal{R}^2$ for the specific choice of the metric (\ref{9}) in this new modified gravity theory. This is because the chosen space-time model is a vacuum solution in general relativity with a negative cosmological constant which remains vacuum in this new gravity theory. However, for non-vacuum solutions with known matter content, the energy-density depends on the coupling constants associated with the higher order terms of the Ricci scalar $\mathcal{R}$. One can demonstrate this by considering examples of such non-vacuum solutions (see, for examples, Refs. \cite{EPJC,NPB}).\\ 

It is noted that the space-time (\ref{9}) is a Misner-like anti-de Sitter space, obtained by identifying points in the covering space of anti-de Sitter space (with radius $\frac{1}{\alpha} = \sqrt{-\frac{3}{\Lambda}}$) related by boost transformations. In suitable coordinates, this Misner-like anti-de Sitter space corresponds to the Lorentzian section of the complex space containing closed time-like curves (CTCs), as constructed in \cite{Li2}. Consequently, the line element (\ref{special-metric}), which also represents a Misner-like anti-de Sitter space, can be understood as the Lorentzian section of the complex space with CTCs as described by
\begin{equation}
    ds^2=(dw-w\,d\phi)^2+\frac{1}{\alpha^2_{*}}\,d\phi^2+(dy-y\,d\phi)^2+(dz-z\,d\phi)^2,\label{special-metric2}
\end{equation}
where $\frac{1}{\alpha_*}=\sqrt{-\frac{3}{\Lambda_m}}=\sqrt{-\left(\frac{\Lambda}{3}+\frac{\beta}{\Lambda}+\frac{4}{3\,\Lambda^2}\,(\gamma+4\,\delta)\right)^{-1}}$ is the new radius of anti-de Sitter space.\\ 

According to quantum field theory (QFT), empty space is defined by the vacuum state, which consists of a collection of quantum fields. These fields exhibit fluctuations even in their ground state (the lowest energy state), due to the presence of zero-point energy, which permeates all of space. In general relativity (GR), the cosmological constant $\Lambda$ is interpreted as an intrinsic energy density of the vacuum, $\rho_{vac}$, with an associated pressure. Based on cosmological observations, $\Lambda$ in GR is found to have a value on the order of $10^{-122}\,\ell^{-2}_{p}$, where $\ell_{p}$ is the Planck length.\\ 

When applying QFT to a new space-time background, such as the one defined by the metric (\ref{special-metric}), the value of an effective cosmological constant $\Lambda_m$ can be determined by selecting appropriate values for the coupling constants $\beta,\gamma,\delta$. These constants influence the behavior of the quantum fields and, consequently, the vacuum energy density within this modified gravity framework.\\

Generally, quantum field theory in space-times with closed timelike curves (CTCs) is not well-defined. However, in the case of Misner space, a well-defined quantum field theory can be constructed in its covering space, which is Minkowski space. In Minkowski space, there are two familiar vacuum states: the Minkowski vacuum, which is invariant under the Poincaré group, and the Rindler vacuum, which is invariant under Lorentz boost transformations. For a detailed discussion of this topic, refer to \cite{Li3}, where the authors identified a self-consistent vacuum state for a massless, conformally coupled scalar field, termed the "adapted" Rindler vacuum. Additionally, in \cite{LX}, it was shown that a self-consistent vacuum also exists for a massless, conformally coupled scalar field in the Misner-like anti-de Sitter space. Thus, we believe it is possible to demonstrate the existence of a self-consistent vacuum state for a massless, conformally coupled scalar field in the Misner-like anti-de Sitter space solution (\ref{special-metric2}) within the context of modified gravity.

\section{The Equations of Motion: Comparison of GR and Ricci-inverse gravity results}\label{sec:4}

In this section, we now study the geodesic motions of test particles in the space-time background given by (\ref{special-metric}) and compare the results with those obtained in general relativity (GR).

The geodesics equations are given by
\begin{equation}
    \ddot{x}^{\lambda}+\Gamma^{\lambda}_{\mu\nu}\,\dot{x}^{\mu}\,\dot{x}^{\nu}=0,\label{B15}
\end{equation}
where dot represents derivative w. r. t. affine parameter $\tau$.

For the metric (\ref{special-metric}), we finds
\begin{eqnarray}
    \frac{\ddot{t}}{t}&=&\frac{\Lambda_m}{3}\,\Big[(\dot{t}-t\,\dot{\phi})^2-(\dot{y}-y\,\dot{\phi})^2-(\dot{z}-z\,\dot{\phi})^2\Big]+\dot{\phi}^2,\label{B16}\\
    \ddot{\phi}&=&\frac{\Lambda_m}{3}\,\Big[(\dot{t}-t\,\dot{\phi})^2-(\dot{y}-y\,\dot{\phi})^2-(\dot{z}-z\,\dot{\phi})^2\Big],\label{B17}\\
    \frac{\ddot{y}}{y}&=&\frac{\Lambda_m}{3}\,\Big[(\dot{t}-t\,\dot{\phi})^2-(\dot{y}-y\,\dot{\phi})^2-(\dot{z}-z\,\dot{\phi})^2\Big]+\dot{\phi}^2,\label{B18}\\
    \frac{\ddot{z}}{z}&=&\frac{\Lambda_m}{3}\,\Big[(\dot{t}-t\,\dot{\phi})^2-(\dot{y}-y\,\dot{\phi})^2-(\dot{z}-z\,\dot{\phi})^2\Big]+\dot{\phi}^2.\label{B19}
\end{eqnarray}

The metric tensor $g_{\mu\nu}$ for the space-time (\ref{special-metric}) depends on the coordinates $t, y$, and $z$, and is independent of the coordinate $\phi$. Therefore, there exist a Killing vector $\partial_{\phi}=\frac{\partial}{\partial \phi}$. The corresponding constant of motion with respect to the parameter $\tau$ can be derived using the relation $k=g_{\mu\nu}\,\xi^{\mu}\,\frac{dx^{\nu}}{d\tau}$. This constant $(\mathrm{L}_{\phi})$ is given by
\begin{equation}
    \mathrm{L}_{\phi}=\left(-\frac{3}{\Lambda_m}+y^2+z^2-t^2\right)\,\dot{\phi}-(y\,\dot{y}+z\,\dot{z}-t\,\dot{t}).\label{killing}
\end{equation}

For a Zero Angular Momentum Observer (ZAMO), the angular momentum, $\mathrm{L}_{\phi}=0$, which indicates that the local inertial reference frame is freely falling and non-rotating, having zero angular momentum. Thus, from Eq. (\ref{killing}), we find
\begin{eqnarray}
    Y\,\frac{d\phi}{d\tau}=\frac{1}{2}\,\frac{dY}{d\tau}\Rightarrow
    \phi=\frac{1}{2}\,\mbox{ln}\left(-\frac{3}{\Lambda_m}+y^2+z^2-t^2\right),\label{killing2}
\end{eqnarray}
where $Y=-\frac{3}{\Lambda_m}+y^2+z^2-t^2$.\\

Hence, the inertial frames have an angular velocity given by
\begin{equation}
    \omega_{RIG}=\frac{\dot{\phi}}{\dot{t}}=\frac{d\phi}{dt}=-\frac{t}{\left(-\frac{3}{\Lambda_m}+y^2+z^2-t^2\right)}\label{killing6}
\end{equation}
which depends on the coupling constants $\beta, \gamma$, and $\delta$ of the modified gravity. Whereas in GR, this result will be
\begin{equation}
    \omega_{GR}=-\frac{t}{\left(-\frac{3}{\Lambda}+y^2+z^2-t^2\right)}.\label{killing7}
\end{equation}

From Eqs. (\ref{killing6})--(\ref{killing7}), we observe that the angular velocity of local inertial frame in modified gravity, $\omega_{RIG}$ is more than that of  $\omega_{GR}$, the corresponding result in general relativity. This increase in the angular velocity is due to the presence of the coupling constants $\beta, \gamma$, and $\delta$ in the modified gravity.\\

The norm of the spacelike killing vector $\partial_{\phi}$ is given by
\begin{equation}
    {\bf X}_{RIG}=||\partial_{\phi}||^2=|g(\partial_{\phi},\partial_{\phi})|=|g_{\phi\phi}|=\left|-\frac{3}{\Lambda_m}+y^2+z^2-t^2\right|\label{killing3}
\end{equation}
which depends on the coupling constants $\beta, \gamma$, and $\delta$ of the modified gravity. The result reduces to the general relativity one (${\bf X}_{RIG} \to {\bf X}_{GR}$) for $\beta=0=\gamma=\delta$. This Killing vector becomes null on the hypersurface $t^2=\left(-\frac{3}{\Lambda_m}+y^2+z^2\right)$.\\

The arc length for closed time-like curve $\bar{\gamma}$ defined by $\Big\{\bar{\gamma}: y=y_0,\, z=z_0,\, \phi \sim \phi+2\,n\,\pi,\quad t_{*}>t_{GR}\,\left(=\sqrt{-\frac{3}{\Lambda}+y^2_{0}+z^2_{0}}\right)\Big\}$ in general relativity is given by 
\begin{equation}
    \ell_{\gamma}=\int\,\sqrt{-g_{\phi\phi}}\,d\phi=2\,|n|\,\pi\,\sqrt{-g_{\phi\phi}}=2\,|n|\,\pi\,\sqrt{t^2_{*}-t^2_{GR}},\label{killing4}
\end{equation}
Whereas in Ricci-inverse gravity, this arc length will be
\begin{equation}
    \tilde{\ell}_{\gamma}=2\,|n|\,\pi\,\sqrt{t^2_{**}-t^2_{RIG}},\label{killing5}
\end{equation}
where the time-like region is defined by $t_{**}>t_{RIG}\,\left(=\sqrt{-\frac{3}{\Lambda_m}+y^2_{0}+z^2_{0}}\right)$. Since $t_{GR}<t_{RIG}$, therefore, $t_{*}$ is smaller than $t_{**}$, that is, $t_{*}<t_{**}$.\\

\begin{figure}
    \centering
    \includegraphics[width=0.35\linewidth]{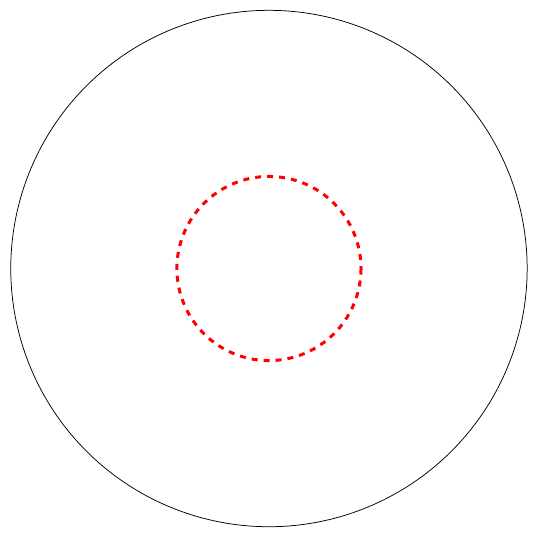}
    \caption{A comparison of the circumference Eqs. (\ref{killing4}) \& (\ref{killing5}) in GR and Ricci-inverse gravity. Here, we set $y_0=0.01,\,z_0=0.01,\,\Lambda=-0.9,\,\beta=0.5,\,\gamma=0.1,\,\delta=0.1$, $t_{*}=2$, and $t_{**}=6$. The dotted red circle represents the result in general relativity and the black one in modified gravity.}
    \label{fig:0}
\end{figure}

From Eqs. (\ref{killing4})--(\ref{killing5}), we observe that the length of the closed time-like curve in modified gravity, $\tilde{\ell}_{\gamma}$ is much more than  $\ell_{\gamma}$, the corresponding result in general relativity. This increase in length in modified gravity is due to the presence of the coupling constants $\beta, \gamma$, and $\delta$. The factor $2\,|n|\,\pi$ arises from the periodicity of the topological structure in the space-time geometry. Taking $r_{*}=\sqrt{t^2_{*}-t^2_{GR}}$ and $r_{**}=\sqrt{-g_{\phi\phi}}=\sqrt{t^2_{**}-t^2_{RIG}}$ as the radius of these arc lengths (now we called circumferences), we depict these in Figure \ref{fig:0}.\\

Now, from geodesic equations (\ref{B16})--(\ref{B19}), we finds the following relation
\begin{equation}
    \frac{\ddot{t}}{t}=\frac{\ddot{y}}{y}=\frac{\ddot{z}}{z}=\ddot{\phi}+\dot{\phi}^2.\label{B20}
\end{equation}

For time-like or null geodesics, using the metric (\ref{special-metric}), we obtain the Lagrangian as follows:
\begin{equation}
    \Big[-(\dot{t}-t\,\dot{\phi})^2-\frac{3}{\Lambda_m}\,\dot{\phi}^2+(\dot{y}-y\,\dot{\phi})^2+(\dot{z}-z\,\dot{\phi})^2\Big]=2\,\mathcal{L}=\epsilon,\label{lag}
\end{equation}
where $\epsilon=0$ for null geodesics and $\epsilon=-1$ for time-like geodesics. 

\begin{figure}
    \centering
    \includegraphics[width=0.45\linewidth]{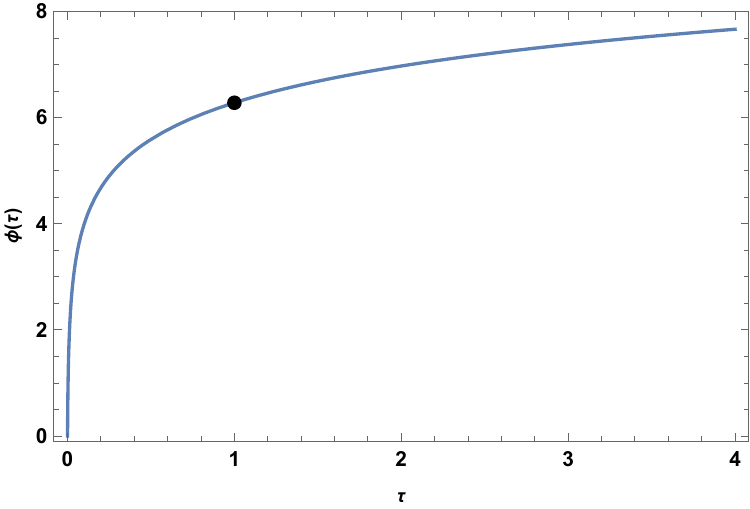}
    \caption{The angular coordinate $\phi$ as a function of $\tau$. Here $\tau_0=1$ and $n=1$.}
    \label{fig:1}
\end{figure}

For null geodesics, from Eqs. (\ref{B17}) and (\ref{lag}) one will finds 
\begin{equation}
    \ddot{\phi}+\dot{\phi}^2=0\label{mm1}
\end{equation}
whose solution is given by 
\begin{equation}
    \phi(\tau)=\mbox{ln}\left|\frac{\tau}{\mathrm{a}}-1\right|,\quad\quad \mathrm{a}=\frac{\tau_0}{1+e^{2\,n\pi}},\label{mm2}
\end{equation}
where we imposed the boundary condition that $\phi(\tau=0)=\phi_0=0$ and $\phi(\tau=\tau_0)=2\,n\,\pi$ since $\phi$ is a periodic coordinate. We have generated Figure \ref{fig:1} showing $\phi$ as a function of $\tau$.\\

Hence, other relations in Eq. (\ref{B20}) also vanishes, that is, $\frac{\ddot{t}}{t}=\frac{\ddot{y}}{y}=\frac{\ddot{z}}{z}=0$. This relation implies the following null geodesic paths given by 
\begin{eqnarray}
    &&t(\tau)=c_1\,\tau+c_2,\nonumber\\
    &&y(\tau)=c_3\,\tau+c_4,\nonumber\\ 
    &&z(\tau)=c_5\,\tau+c_6,\label{lag4}
\end{eqnarray}
where $c_i\,(i=1,\ldots ,6)$ are arbitrary constants. Thus, we conclude that the null geodesics motions remain same both in Ricci-inverse gravity and general relativity theories. However, this null geodesics paths are not closed, $x^{\mu}(\tau=0) \neq x^{\mu}(\tau=\tau_0)$, and hence, no closed null geodesics will be formed. \\

Now, we focus on time-like geodesics, that is, motions of massive test particles in this new space-time background (\ref{special-metric}). In that case, from Eqs. (\ref{B17}) and (\ref{lag}) we finds
\begin{equation}
    \ddot{\phi}+\dot{\phi}^2=\frac{\Lambda_m}{3}=-\mathrm{b}^2,\label{lag2}
\end{equation}
where $\mathrm{b}$ is given by
\begin{equation}
    \mathrm{b}=\sqrt{-\frac{\Lambda_m}{3}}=\sqrt{-\frac{\left(\Lambda+\frac{3\,\beta}{\Lambda}+\frac{4\,\gamma}{\Lambda^2}+\frac{16\,\delta}{\Lambda^2}\right)}{3}}.\label{lag3}
\end{equation}

Solving equation (\ref{lag2}) with the boundary conditions that $\phi(\lambda=0)=\phi_0=0$ and $\phi(\lambda=\lambda_0)=2\,n\,\pi$ since $\phi$ is a periodic coordinate, we finds
\begin{equation}
    \phi(\lambda)=\frac{1}{\mathrm{b}}\,\mbox{ln} \left|\frac{\cos C}{\cos (\mathrm{b}\,\lambda-C)}\right|,\label{B21}
\end{equation}
where the constant $C$ is to be determined by solving the following equation:
\begin{equation}
   \cos (\mathrm{b}\,\lambda_0-C)=e^{-2\,n\,\mathrm{b}\,\pi}\,\cos C\quad \mbox{but}\quad b\,\lambda_0 \neq 2\,n\,\pi.\label{B22}
\end{equation}

Similarly, other equations with their solutions for the time-like geodesics case using (\ref{B20}) and (\ref{lag2}) are as follows:
\begin{eqnarray}
    &&\ddot{t}+\mathrm{b}^2\,t=0\Rightarrow t(\lambda)=\mathcal{A}_1\,\cos (\mathrm{b}\,\lambda)+\mathcal{B}_1\,\sin (\mathrm{b}\,\lambda),\label{B23}\\
    &&\ddot{y}+\mathrm{b}^2\,y=0\Rightarrow y(\lambda)=\mathcal{A}_2\,\cos (\mathrm{b}\,\lambda)+\mathcal{B}_2\,\sin (\mathrm{b}\,\lambda),\label{B24}\\
    &&\ddot{z}+\mathrm{b}^2\,z=0 \Rightarrow z(\lambda)=\mathcal{A}_3\,\cos (\mathrm{b}\,\lambda)+\mathcal{B}_3\,\sin (\mathrm{b}\,\lambda),\label{B25}
\end{eqnarray}
where $\mathcal{A}_i, \mathcal{B}_i$ are arbitrary constants and and $\mathrm{b}$ is given in Eq. (\ref{lag3}). The above paths are periodic function with the periodicity $T_{RIG}$  given by 
\begin{equation}\label{periodic1}
    T_{RIG}=\frac{2\,\pi}{\mathrm{b}}=2\,\pi\,\sqrt{-\frac{3}{\left(\Lambda+\frac{3\,\beta}{\Lambda}+\frac{4\,\gamma}{\Lambda^2}+\frac{16\,\delta}{\Lambda^2}\right)}}.
\end{equation}
Whereas, in general relativity, it is as follows:
\begin{equation}\label{periodic2}
    T_{RG}=2\,\pi\,\sqrt{-\frac{3}{\Lambda}},
\end{equation}
with the time-like geodesics paths given by
\begin{eqnarray}
    &&t(\lambda)=\mathcal{C}_1\,\cos \left(\sqrt{-\frac{\Lambda}{3}}\,\lambda\right)+\mathcal{D}_1\,\sin \left(\sqrt{-\frac{\Lambda}{3}}\,\lambda\right),\label{B26}\\
    &&y(\lambda)=\mathcal{C}_2\,\cos \left(\sqrt{-\frac{\Lambda}{3}}\,\lambda\right)+\mathcal{D}_2\,\sin \left(\sqrt{-\frac{\Lambda}{3}}\,\lambda\right),\label{B27}\\
    &&z(\lambda)=\mathcal{C}_3\,\cos \left(\sqrt{-\frac{\Lambda}{3}}\,\lambda\right)+\mathcal{D}_3\,\sin \left(\sqrt{-\frac{\Lambda}{3}}\,\lambda\right).\label{B28}
\end{eqnarray}

From the above time-like geodesics analysis  both in general relativity and Ricci-inverse gravity theories, we see that geodesics paths of massive test particles in this new space-time background (\ref{special-metric}) is sinusoidal varying modes. Moreover, we see that the time-like geodesic paths depend on the coupling constants $\beta,\,\gamma,\,\delta$. All these results reduces to the general relativity one for zero coupling constants $\beta=0=\gamma=\delta$. However, this time-like geodesics paths are not closed, $x^{\mu}(\lambda=0) \neq x^{\mu}(\lambda=\lambda_0)$, and hence, no closed time-like geodesics will be formed.\\

\begin{figure}[ht!]
\begin{centering} 
\includegraphics[width=0.55\linewidth]{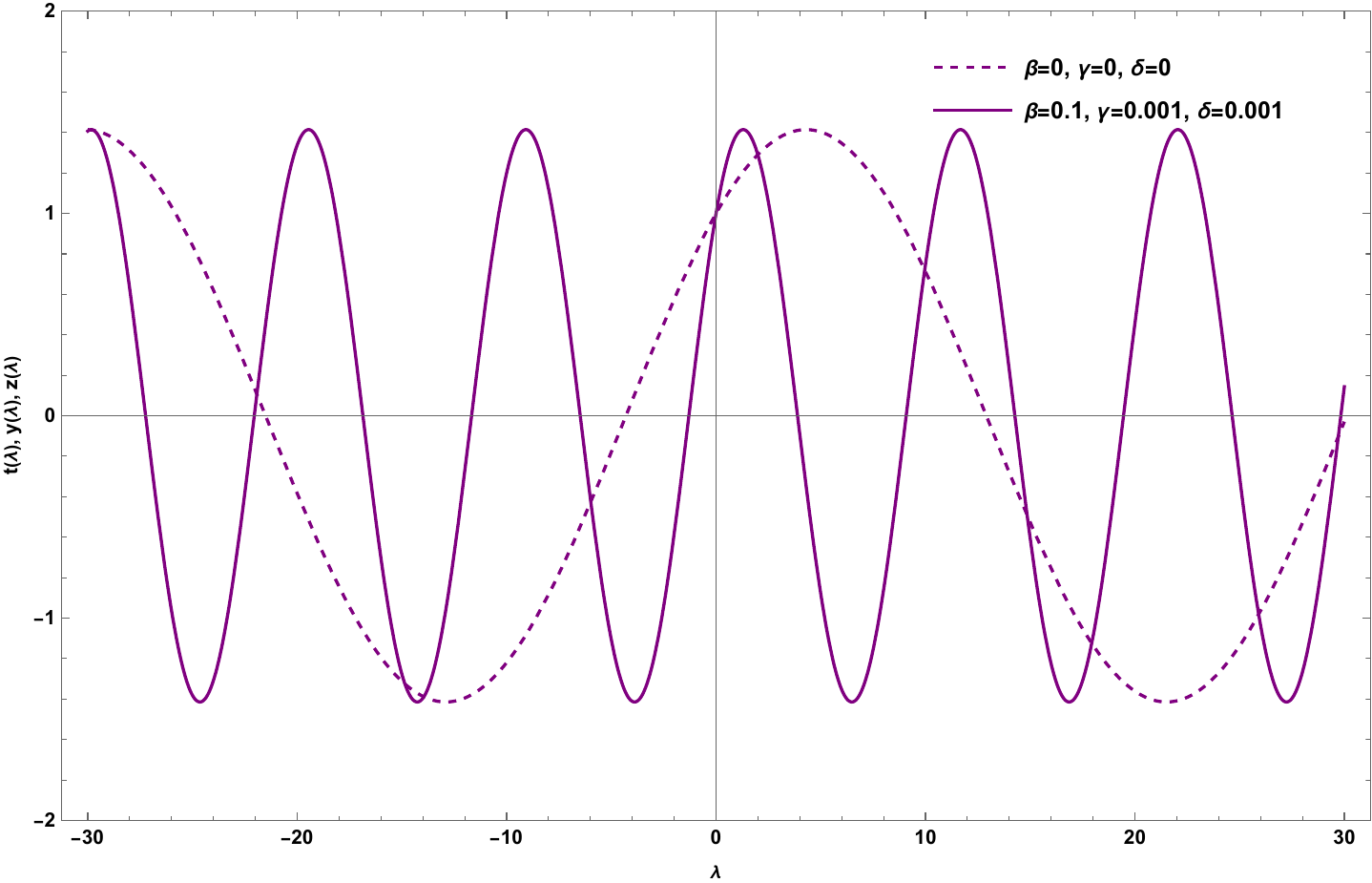}
\caption{Time-like Geodesics paths $t(\lambda), y(\lambda), z(\lambda)$ for massive objects. Here $\mathcal{A}_i=1=\mathcal{B}_i$ and $\Lambda=-0.1$. Solid line represents result in Ricci-inverse gravity and dotted line in GR.}
\label{fig:2}
\hfill\\
\includegraphics[width=0.55\linewidth]{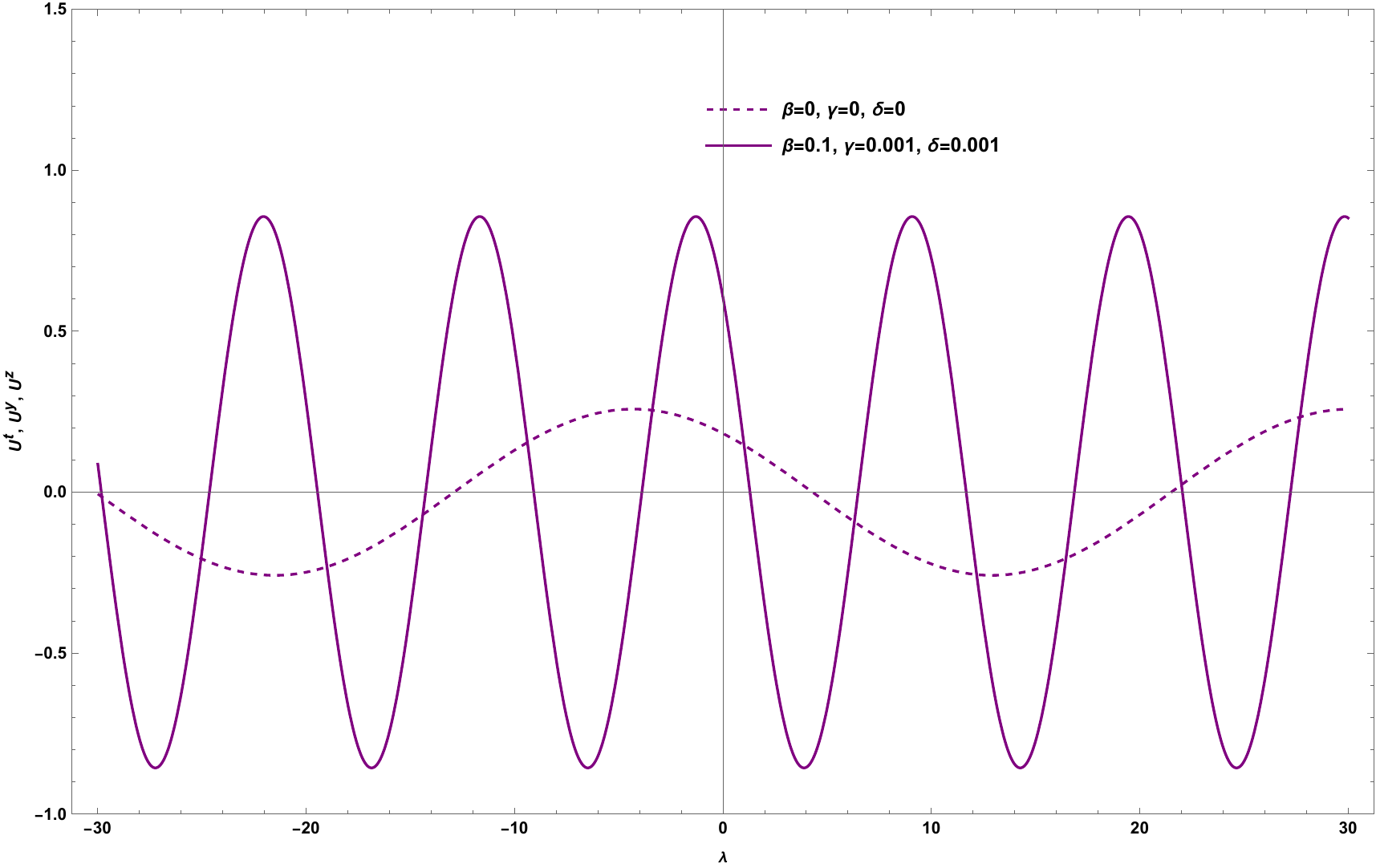}
\caption{Velocities $U^t, U^y, U^z$ for massive objects. Here $\mathcal{A}_i=1=\mathcal{B}_i$ and $\Lambda=-0.1$. Solid line represents result in Ricci-inverse gravity case and dotted line in GR.}
\label{fig:3}
\end{centering}
\end{figure}

Another interesting point we have observed is that the time-period ($T_{RIG}$) of this oscillatory motion is lesser in Ricci-inverse gravity compared to general relativity case (see Figure \ref{fig:2}). This indicates that massive objects moving faster in Ricci-inverse gravity background compared to general relativity case.  To reinforce this result, let us calculate the velocity components $U^{t}$, $U^{y}$, and $U^{z}$ of a massive object as functions of $\lambda$. These quantities are defined by
\begin{equation}
    U^{\mu}=\frac{dx^{\mu}}{d\lambda}.\label{BB26}
\end{equation}

Using equations (\ref{B21}), and (\ref{B23})--(\ref{B25}), we finds
\begin{eqnarray}
    &&U^{t}=\mathrm{b}\,\Big[-\mathcal{A}_1\,\sin (\mathrm{b}\,\lambda)+\mathcal{B}_1\,\cos (\mathrm{b}\,\lambda)\Big],\label{BB27}\\
    &&U^{y}=\mathrm{b}\,\Big[-\mathcal{A}_2\,\sin (\mathrm{b}\,\lambda)+\mathcal{B}_2\,\cos (\mathrm{b}\,\lambda)\Big],\label{BB29}\\
    &&U^{z}=\mathrm{b}\,\Big[-\mathcal{A}_3\,\sin (\mathrm{b}\,\lambda)+\mathcal{B}_3\,\cos (\mathrm{b}\,\lambda)\Big].\label{BB30}
\end{eqnarray}

In Figure \ref{fig:3}, we have plotted the components $U^{t}$, $U^{y}$, and $U^{z}$ as functions of $\lambda$ for both Ricci-inverse gravity and general relativity. We observe that the behavior is similar to what is shown in Figure \ref{fig:2}.

\section{Time-machines construct within $f(\mathcal{R})$-gravity theory}\label{sec:5}

In this section, we analyze the same time-machine model represented by (\ref{9}) within the framework of a modified gravity theory known as $f(\mathcal{R})$-gravity. In this theory, the Ricci scalar $\mathcal{R}$ is replaced by a function of itself, $\mathcal{R} \to f(\mathcal{R})$, in the Einstein-Hilbert action. As noted in Section \ref{sec:3}, we found that the $\mathcal{R}^2$ term did not affect the results derived in that section. Therefore, we will consider a polynomial function of
$\mathcal{R}$ with an order $n>2$.\\

Therefore, the Lagrangian that describes this $f(\mathcal{R})$-gravity theory is given by
\begin{eqnarray}
    S= \int d^4x\, \sqrt{-g}\left[f(\mathcal{R})-2\,\Lambda_m+{\cal L}_m\right],\label{ss1}
\end{eqnarray}
where $\Lambda _m$ is an effective cosmological constant. Varying the action with respect to the metric tensor results the modified field equations for $f(\mathcal{R})$-gravity with the cosmological constant given by 
\begin{equation}\label{ss2}
    f_{\mathcal{R}}\,R_{\mu \nu}-\frac{1}{2}\, f(\mathcal{R})\, g_{\mu \nu}-\nabla _{\mu}\, \nabla _{\nu} \, f_{\mathcal{R}} + g_{\mu \nu}\, \nabla _{\alpha}\, \nabla ^{\alpha} \, f_{\mathcal{R}} + \Lambda _m\, g_{\mu \nu}=\mathcal{T}_{\mu \nu}\, 
\end{equation}
where we have defined
\begin{equation}\label{ss3}
    f_{\mathcal{R}} = \frac{\partial f(\mathcal{R})}{\partial \mathcal{R}}\, .
\end{equation}

Now, we choose the function $f(\mathcal{R})$ to be the following form 
\begin{equation}\label{ss4}
    f(\mathcal{R})=\mathcal{R}+a_n \,\mathcal{R}^{n},\quad n > 2.
\end{equation}
The reason for $n>2$ is obvious because we already have considered a term $\mathcal{R}^2$ in the function $f(\mathcal{R},\mathcal{A},A^{\mu\nu}\,A_{\mu\nu})$-gravity theory discussed in section \ref{III3} and showed that there is no effects on the results due to this specific term. However, maybe there is effects from higher order terms which we discuss below. 

Thus, we finds
\begin{equation}\label{ss5}
    f_{\mathcal{R}}=1+a_n \,n\,\mathcal{R}^{n-1}.
\end{equation}
It is important to note that, considering the time-machine solution (\ref{Li}), we have that 
\begin{equation}\label{condition}
-\nabla _{\mu}\, \nabla _{\nu} \, f_{\mathcal{R}} + g_{\mu \nu}\, \nabla _{\alpha}\, \nabla ^{\alpha}\,f_{\mathcal{R}} =0     
\end{equation}
since the Ricci scalar $\mathcal{R}=4\,\Lambda$ is a constant.

Substituting (\ref{ss4})--(\ref{ss5}) into the modified field equations (\ref{ss2}), we finds
\begin{equation}\label{ss6}
(1+a_n \,n\,\mathcal{R}^{n-1})\,R_{\mu \nu}-\frac{1}{2}\,(\mathcal{R}+a_n \,\mathcal{R}^{n})\, g_{\mu \nu}+\Lambda _m\, g_{\mu \nu}=\mathcal{T}_{\mu \nu}\, .
\end{equation}

Using the Ricci tensor $R_{\mu\nu}$ and the Ricci scalar $\mathcal{R}$ given in Eq. (\ref{7a}), we can rewrite the above field equations (\ref{ss6}) as follows:
\begin{equation}\label{ss7}
    \left(\Lambda_m-\Lambda+n\,a_n\,\Lambda^n\,2^{2\,n-2}-a_n\,\Lambda^n\,2^{2\,n-1}\right)\,g_{\mu\nu}=\mathcal{T}_{\mu \nu}.
\end{equation}
Solving the modified field equations (\ref{ss7}) for vacuum with the energy-momentum tensor $\mathcal{T}_{\mu \nu}=0$, and after simplification, we finds
\begin{equation}\label{ss8}
    \Lambda _m = \Lambda +2^{2\,n-2}\,(2-n)\,a_n\,\Lambda^n\, .
\end{equation}
Thus, we confirm that the time-machine solution (\ref{9}) constructed in GR is also a vacuum solution in $f(\mathcal{R})$-gravity theory with an effective cosmological constant $\Lambda_m$ given in Eq. (\ref{ss8}). Hence, formation of closed time-like curves is possible in this $f(\mathcal{R})$-gravity theory also. Noted here that for $n=2$, from Eq. (\ref{ss8}) we retains the general relativity result $\Lambda_m \to \Lambda$. This means that there is no effects of only the term $\mathcal{R}^2$ but have for the case $n >2$.\\

Therefore, the vacuum space-time (\ref{9}) in $f(\mathcal{R})$-gravity theory with the function $f(\mathcal{R})=\mathcal{R}+a_n\,\mathcal{R}^n$ can be described by the following line-element  
\begin{equation}
    ds^2=-(dt-t\,d\phi)^2+\frac{3}{(-\Lambda_m)}\,d\phi^2+(dy-y\,d\phi)^2+(dz-z\,d\phi)^2,\label{special-metric3}
\end{equation}
where $\Lambda_m$ is given in Eq. (\ref{ss8}).\\

Analogue to the previous analysis done in section \ref{sec:4}, we calculate the following physical quantities and present the results. The angular velocity of the local inertial frame in $f(\mathcal{R})$-gravity is given by
\begin{equation}
    \omega_{f(\mathcal{R})}=-\frac{t}{\left[-\frac{3}{\left(\Lambda+2^{2\,n-2}\,(2-n)\,a_n\,\Lambda^n\right)}+y^2+z^2-t^2\right]}.\label{killing9}
\end{equation}
The norm of the spacelike killing vector $\partial_{\phi}$ is given by
\begin{equation}
    {\bf X}_{f(\mathcal{R})}=\left|-\frac{3}{\Lambda+2^{2\,n-2}\,(2-n)\,a_n\,\Lambda^n}+y^2+z^2-t^2\right|.\label{killing10}
\end{equation}
And the arc length of closed time-like curve for the closed curve $\bar{\bar\gamma}$ defined by $\Big\{\bar{\bar\gamma}:\, y=y_0,\,\,z=z_0,\,\,\phi \sim \phi+2\,n\,\pi,\,\, t=t_{***}>t_{f(\mathcal{R})}\left(=\sqrt{-\frac{3}{[\Lambda +2^{2\,n-2}\,(2-n)\,a_n\,\Lambda^n]}+y^2_{0}+z^2_{0}}\right) \Big\}$ will be
\begin{equation}
    \bar{\bar\ell}_{\gamma}=2\,|n|\,\pi\,\sqrt{t^2_{***}-t^2_{f(\mathcal{R})}}.\label{killing11}
\end{equation}
A comparison of results in $f(\mathcal{R})$-gravity with general relativity and Ricci-inverse gravity can be made for a specific value of $n$.\\

Below, we present a few example of this $f(\mathcal{R})$-gravity theory. The simplest example of this function $f(\mathcal{R})$ is given by
\begin{equation}\label{ss9}
    f(\mathcal{R})=\mathcal{R}+a_3\,\mathcal{R}^3,\quad a_3>0\,.
\end{equation}
Thus, from Eq. (\ref{ss8}), we finds an effective cosmological constant given by
\begin{equation}\label{ss10}
    \Lambda _m = \Lambda\,(1-2^{4}\,a_3\,\Lambda^2)<\Lambda,
\end{equation}
provided we have the constraint on the coefficient $a_3 < \frac{1}{2^4\,\Lambda^2}$.\\

Next, we consider another simplest example of function $f(\mathcal{R})$ given by
\begin{equation}\label{ss11}
    f(\mathcal{R})=\mathcal{R}+a_3\,\mathcal{R}^3+a_4\,\mathcal{R}^4,\quad a_3>0,a_4>0\,.
\end{equation}

Thereby, substituting this function (\ref{ss11}) and using (\ref{condition}) into the modified field equations (\ref{ss2}), and after solving for vacuum as the matter-content, we finds an effective cosmological constant given by
\begin{equation}\label{ss12}
    \Lambda _m = \Lambda\,(1-16\,a_3\,\Lambda^2-192\,a_4\,\Lambda)<0,
\end{equation}
provided we have the constraint on the coefficient 
\begin{equation}\label{ss13}
    a_3 < \left(\frac{1}{2^4\,\Lambda^2}-12\,a_4\,\Lambda\right),\quad a_4>0,\quad \Lambda<0\,.
\end{equation}

Now, considering a general function of $f(\mathcal{R})$ in the following form 
\begin{equation}\label{ss14}
f(R)=\mathcal{R}+a_3\,\mathcal{R}^3+a_4\,\mathcal{R}^4+a_5\,\mathcal{R}^5+....+a_n\,\mathcal{R}^n=\mathcal{R}+\sum_{i=3}^{n} a_i\,\mathcal{R}^i,
\end{equation}
where $a_3>0,a_4>0,.....,a_n>0$ are positive coupling constants. It can be shown that the specific time-machine space-time, given by (\ref{9}), constitutes a valid solution within the framework of $f(R)$-modified gravity theory, where the vacuum serves as the energy-momentum tensor and the cosmological constant is negative, with its value dependent on the coupling constants $a_i$. Consequently, the formation of closed time-like curves is feasible in $f(\mathcal{R})$-gravity, including the Ricci-inverse gravity theory.

\section{Conclusions}\label{sec:6}

Standard physics is founded on the principle that causality must not be violated. However, General Relativity permits several exact solutions that allow the formation of Closed Time-like Curves (CTCs), Closed Null Geodesics (CNGs), and Closed Time-like Geodesics (CTGs), leading to potential violations of the concept of causality. In other words, Einstein’s theory of gravitation theoretically allows for the possibility of time machines. Numerous solutions to Einstein’s field equations have been constructed under various matter distributions, both with and without the cosmological constant, in the framework of General Relativity. Notably, most of these solutions admit circular CTCs or CTGs at certain values of radial distance, $r$.\\

A different category of CTC models exists in which CTCs emerge instantaneously from an initially spacelike hypersurface in a causally well-behaved manner. The first such model is the two-dimensional Misner space, where the metric is given by $ds^2_{2D}=-T\,d\psi^2-dT\,d\psi$ \cite{CWM} with $-\infty < T < \infty$ and $\psi$ being a periodic coordinate. Inspired by this model, Li constructed a four-dimensional generalization of the two-dimensional Misner space into curved space-time \cite{LX}. This space-time is a vacuum solution to Einstein’s field equations with a negative cosmological constant and serves as a model for a Time-Machine, featuring CTCs within its temporal region. Further studies showed that a Misner-like anti-de Sitter space could be derived from the covering space of anti-de Sitter space. This Misner-like space contains CTCs, with regions of CTCs separated from non-CTC regions by chronology horizons. Following this, Ori developed a four-dimensional generalization of Misner space in curved space-time, which is also a vacuum solution to the field equations of General Relativity \cite{Ori}. Subsequently, a number of four-dimensional curved space-times, representing generalizations of the two-dimensional Misner space, have been presented in the literature (see, for example, \cite{FA1,FA22,FA3,FA4} and related references therein).\\

In this manuscript, we examined Li's Time-Machine space-time within the framework of a modified gravity theory known as Ricci-inverse gravity. This novel gravitational theory is characterized by the inclusion of an anti-curvature tensor in the Einstein-Hilbert action. In Section \ref{sec:2}, we revisit Li's space-time within the context of General Relativity. In Section \ref{sec:3}, we derived the modified field equations within the framework of Ricci-inverse gravity, using Li's Time-Machine space-time as the background model. We categorize and presented our results in Class-{\bf I} to Class-{\bf III}. Our findings indicated that the background Time-Machine model serves as a solution in all classes of Ricci-inverse gravity theory, with a zero energy-momentum tensor and an effective cosmological constant. While in General Relativity the chosen space-time is a vacuum solution with a negative cosmological constant, we also find that in Ricci-inverse gravity, the effective cosmological constant, as derived in Eqs. (\ref{21}), (\ref{A12}), and (\ref{B14}), remains negative. These negative values of $\Lambda_m$ are achieved by adjusting the coupling constants associated with each class. Thus, we concluded that Li's Time-Machine model, at least theoretically, represents a viable Time-Machine model in Ricci-inverse gravity.\\

In Section \ref{sec:4}, we discussed the equations of motion and present results for both null and time-like geodesics. It is demonstrated that time-like geodesic paths are periodic, with the periodicity $T_{RIG}<T_{GR}$, indicating that massive test particles move faster in Ricci-inverse gravity compared to General Relativity. Furthermore, we provided physical interpretations of the model (\ref{special-metric}), constructed within the framework of Ricci-inverse gravity theory. Notably, we have shown that arc length of the Closed Time-like Curve is greater in this modified gravitational theory compared to General Relativity.\\

In Section \ref{sec:5}, we explored another modified gravity theory, known as $f(\mathcal{R})$-gravity theory, using Li's Time-Machine space-time as the background model. We demonstrate that for $f(\mathcal{R})=(\mathcal{R}+a_n\,\mathcal{R}^n)$, where $n>2$, Li's Time-Machine spacetime remains a valid solution in $f(\mathcal{R})$-gravity. This solution features a zero energy-momentum tensor, $\mathcal{T}_{\mu\nu}=0$, and an effective negative cosmological constant as shown in Eq. (\ref{ss8}). Analogous to the analysis in Section \ref{sec:4}, the solution (\ref{special-metric3}) constructed in the gravity theory can be interpreted and compared with the results from General Relativity. Consequently, we concluded that, in addition to Ricci-inverse gravity, Li's Time-Machine model also represents at least, theoretically a Time-Machine solution in $f(\mathcal{R})$-gravity theory.\\

Table \ref{table:1} compares the results of Class-{\bf I} to Class-{\bf III} with those obtained from General Relativity. Additionally, Table \ref{table:2} summarizes various space-times that serve as valid solutions in Ricci-inverse gravity, considering different matter content. In this paper, we explored two specific modified gravity theories: Ricci-inverse gravity and $f(\mathcal{R})$-gravity, using Li's Time-Machine space-time as the background model. We concluded that the Time-Machine model constructed in General Relativity also represents a viable Time-Machine solution in both Ricci-inverse gravity and $f(\mathcal{R})$ gravity theories.\\

In future studies, we plan to explore other alternative gravitational theories using Li's Time-Machine space-time as the background model and analyze the resulting outcomes.

\section*{Data Availability Statement}

No data were generated or analyzed in this paper.

\section*{Conflicts of Interest}

No conflict of interests in this paper.

\section*{Acknowledgments}

F.A. acknowledges the Inter University Centre for Astronomy and Astrophysics (IUCAA), Pune, India for granting visiting associateship. This work by A. F. S. is partially supported by National Council for Scientific and Technological Develo\-pment - CNPq project No. 312406/2023-1. J. C. R. S. thanks CAPES for financial support.

\end{document}